\documentclass{agujournal2019}
\usepackage{url} 
\usepackage{lineno}
\usepackage[inline]{trackchanges} 
\usepackage{soul}
\usepackage{mathptmx}
\usepackage{amsmath}
\usepackage{physics}

\draftfalse

\journalname{~}

\makeatletter
\def\@journalname{}
\def\@papertype{}
\def\ps@drafttitle{}
\makeatother
\begin{document}

%
%


\title{Thermodynamics of Shear Equilibration During Magnetic Reconnection Onset in Mixed-Equilibrium Current Sheets}

\authors{D. S. Payne\affil{1}, M. Swisdak\affil{2}, J.F. Drake\affil{2}, T.C. Li\affil{3}}

\affiliation{1}{Climate and Space Sciences and Engineering, University of Michigan, Ann Arbor, MI}

\affiliation{2}{Institute for Research in Electronics and Applied Physics, University of Maryland, College Park, MD}

\affiliation{3}{Department of Physics and Astronomy, Dartmouth College, Hanover, NH}


\begin{keypoints}
\item Association between local thermodynamic changes and magnetic shear is examined for a reconnecting mixed-equilibrium current sheet
\item After $\beta_e$ in the current sheet exceeds its asymptotic value, magnetic and thermal pressures rapidly approach their asymptotic values  
\item Both species undergo compressive work, but heat density evolution between species differs and may be mediated by $E_\perp$
\end{keypoints}

\begin{abstract}
Magnetic shear across the polarity inversion line (PIL) plays an important role in the explosive nature of reconnection onset and in the equilibration of current sheets, acting as a source of free energy that can enhance or inhibit the onset process under certain conditions. In this study, we use a 2D PIC simulation to examine the local interaction between the reconnection guide field and thermodynamic variables during reconnection onset in a region of initially depleted thermal energy and enhanced magnetic energy in a large guide field background. We identify critical stages of the equilibration process, characterize intervals based on whether the pressure evolution is driven by changes in density or temperature, and discuss what these intervals imply about the evolution of local heat and work density. Finally, we examine power densities associated with electromagnetic field time evolution and electromagnetic energy transfer and compare to those related to thermodynamic changes.
\end{abstract}

\section{\label{sec:level1}Introduction}

The onset of magnetic reconnection remains one of the most active topics of research in space and heliophysics due to its relevance to solar flare energy release and geomagnetic storms.  The variety of plasma environments in which it occurs make it difficult to generalize, though it is typically an explosive process capable of rapidly releasing large amounts energy stored in the plasma upstream of the current sheet.  In the solar corona, one parameter that has been associated with the "switch-on" nature of reconnection onset is the amount of magnetic shear near the polarity inversion line (PIL) \cite{daldorff2018onset,leake2020onset, Klimchuk2023,leake2024onset}, where larger amounts of magnetic shear were associated with a shorter time to onset and a more rapid kinetic energy release during onset.  It should be noted that magnetic "shear" in this context is not referring to the shear angle between the reconnecting fields, but instead refers to the amount of guide field accumulated at the PIL.  It has been suggested that another critical stage predictive of reconnection onset is when the tearing growth timescale and the current sheet driving scale approach each other \cite{leake2024onset}.  The correlation between energy release timescales and magnetic structure is suggestive of a thermodynamic relationship, and reconnection has even been compared to a phase transition \cite{Jara-Almonte2021}.  Even at much smaller scales such as electron-only reconnection in the turbulent magnetosheath, a relationship between the local magnetic curvature evolution and compressive work has been observed \cite{payne2025situ}, again highlighting a mechanistic relationship between local thermodynamics and changes in local magnetic shear.  Current sheets have also been shown to approach mixed equilibria (neither a Harris-sheet nor force-free) via local guide field amplification processes\cite{yoon2023equilibrium}, and reconnection onset has been shown to be sensitive to the background plasma $\beta$ and the local guide field strength \cite{Yoon2024}.   Thus, the amount of magnetic shear (guide field) and the background thermal and magnetic pressures play a critical role in mixed equilibrium selection, the likelihood of reconnection onset, and the amount of energy release during onset.   

Due to their often collisionless nature, space plasmas often require a more nuanced approach to describing thermodynamic behavior due to the highly non-Maxwellian phase space distributions that commonly form in highly magnetized collisionless plasmas. One approach constructs a generalized version of the first law of thermodynamics that accounts for the phase space evolution of populations far from local thermodynamic equilibrium (LTE) and invokes concepts like the infinitesimal evolution of generalized heat per particle \cite{Cassak2023,Barbhuiya2024}.  

A simplified expression for the generalized first law taking into account the time evolution of heat density and work density is provided below

\begin{equation}
\partial_t u = \partial_t Q + \partial_t W 
\label{eqn:firstlaw}
\end{equation}

where $u$ is the internal energy density, $Q$ is the heat density, and $W$ is the local work density (where density is per unit volume, not per particle as is the case for similar terms in \cite{Cassak2023}).  The internal  energy is typically the total thermal energy content of a system, so the internal energy density is then the thermal energy density which is directly proportional to the scalar thermal pressure where $P_{th} = \frac{2u}{3} \propto nT$.  The local time evolution of thermal pressure and internal thermal energy density has contributions from local temperature and local density evolution.  When $T$ is in energy units, we can neglect the Boltzmann constant and express the time derivative of the thermal pressure in a simpler form:  

\begin{equation}
\partial_t P_{th}= n \ \partial_t T  + T \ \partial_t n
\label{eqn:pressurederivative}
\end{equation}

where $P_{th}$ is the thermal pressure, $n$ is the number density, and $T$ is the temperature.  Note that the terms in the derivative on the left hand side of equations \ref{eqn:firstlaw} and \ref{eqn:pressurederivative}, ($u$ and $P_{th}$ respectively) both represent an energy density associated with the local thermal energy of the particles, and therefore are analogous terms that only differ by a constant.  Therefore, the expression in equation \ref{eqn:pressurederivative} is effectively an alternate local expression of the first law of thermodynamics.  We further make the comparison between the two terms on the right hand side of equations \ref{eqn:firstlaw} and \ref{eqn:pressurederivative}, associating changes in work density with with the evolution of particle density (as argued in \cite{Cassak2023}), leaving the final term as representative of the heat density evolution. It is important to note that the relative contributions by the terms in equations \ref{eqn:firstlaw} and \ref{eqn:pressurederivative} are frame-dependent, due to the fact that temporal evolution is dependent on the reference frame of choice.  In the following sections we are most interested in the thermodynamic behavior of an evolving x-line, so all measurements move with the x-line as it drifts along the outflow axis.  

The above power densities concern only local thermodynamic evolution, and thus cannot describe the behavior of local electromagnetic energy transport and transfer. More relevant power densities include the time evolution of the local magnetic and electric pressures ($\partial_t P_M$ and $\partial_t P_E$, respectively) and $\vec{J}\cdot\vec{E}$, which represents the local power density of energy transfer between electromagnetic fields and plasma particles.  Expressed in simplified units which ignore the vacuum permittivity and permeability, the time derivatives of the electric and magnetic pressures are defined below.  

\begin{equation}
\begin{aligned}
\partial_t P_E = \partial_t (E^2/2c) \\
\partial_t P_M = \partial_t (B^2/2)
\label{empressurederivatives}
\end{aligned}
\end{equation}

By comparing the magnitudes and sequence of the electrodynamic power densities with thermodynamic power density contributions, we can get a better grasp on the local magneto-thermodynamic behavior of an x-line as it forms with a significant guide field.

\section{\label{sec:level2}Simulation Setup}

The simulation in this study was run using P3D \cite{zeiler2002three}, a particle-in-cell (PIC) code.  The simulation domain is a 2 dimensional grid of $L_x \times L_y =2048 \times 1024$ grid points equivalent to $51.2 \ d_i \times 25.6 \  d_i$ where $d_i$ is the ion skin depth, and with 400 particles per cell.  The mass and temperature ratios are $\frac{m_i}{m_e} = 25$ and $\frac{T_i}{T_e} = 10$, respectively.  Reconnection is initiated and localized via a small perturbation in the normal  $B_y$ component of the magnetic field.  

The initial configuration of magnetic fields is a force-free magnetic profile with a $B_x$ reversal and with a large guide field $B_z$ that balances it to maintain spatially uniform magnetic pressure $P_M$, with variation only along the y axis.

\begin{equation}
\begin{aligned} 
B_x(y) = \tanh(y/w_0) \\
B_z(y) = \sqrt{5-B^2_x}    
\end{aligned}
\end{equation}

The initial distribution of particles is uniform in temperature ($T$) , but includes a density ($n$) depletion which takes the following form

\begin{equation}
n(y) = 1 + 0.5*\tanh^2(y/w_0)    
\end{equation}

In addition to the initially uniform $P_M$, the $n$ depletion necessarily produces a depletion in the total pressure ($P_{tot}$) at the PIL and the system starts in a pressure-imbalanced state.  Once the simulation time begins, the current sheet is allowed to evolve naturally toward an approximate equilibrium rather than being imposed with one.

\section{\label{sec:level3} Initial Current Sheet Equilibrium}
\begin{figure}[h]
\centering
\includegraphics[scale=0.3]{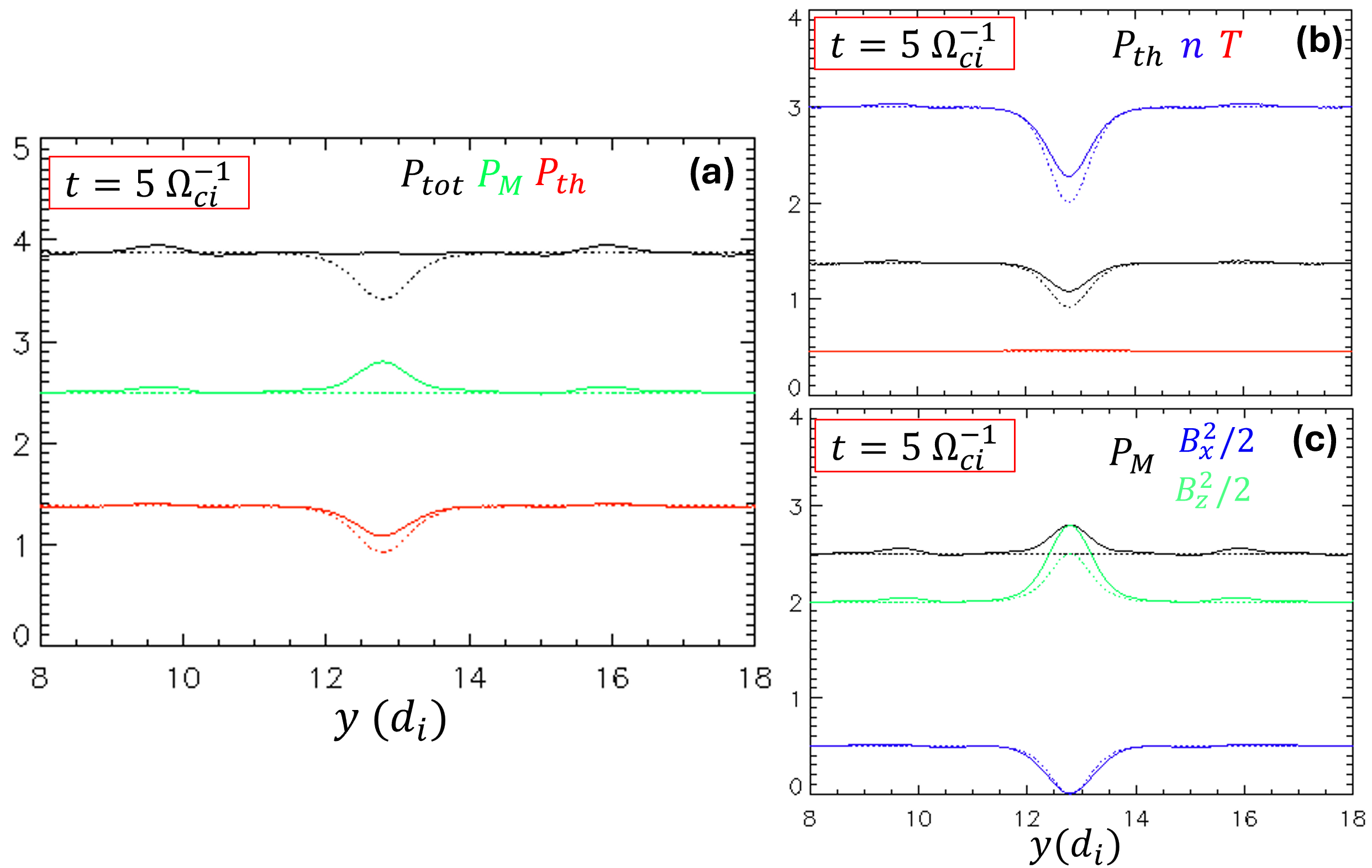}
\caption{Initial Current Sheet Equilibration via Guide Field Amplification.  (a) mean $P_{tot}$, $P_M$, and $P_{th}$ profiles for a cut normal to the current sheet in black, green, and red, respectively.  (b) $P_{th}$, $n$ and $T$ profiles for a cut normal to the current sheet in black, blue, and red, respectively.  (c) $P_M$ and the $B_x$ and $B_y$ contributions to it for a cut normal to the current sheet in black, blue, and green, respectively.  In all panels, solid lines indicate the state of the variables at $t \sim 5 \ \Omega^{-1}_{ci}$ while dotted lines indicate the initial state of the variables.}
\label{Setup}
\end{figure}

The initial configuration has a net pressure depletion due to a depletion in $n$ around the current sheet.  Figure \ref{Setup} shows how within a few  $\Omega_{ci}^{-1}$, the system responds to achieve approximate pressure balance by a simultaneous increase of $n$ and $B_z^2$ the depleted pressure region.  During this brief interval, $\beta$ at the center of the current sheet remains relatively unchanged as the magnetic and thermal pressures ($P_M$ and $P_{th}$, respectively) evolve.  The result is a mixed equilibrium, where the local $P_{th}$ depletion is roughly balanced by the local $P_M$ enhancement.   

\section{\label{sec:level4} The Onset Interval}
\begin{figure}[h]
\centering
\includegraphics[scale=0.4]{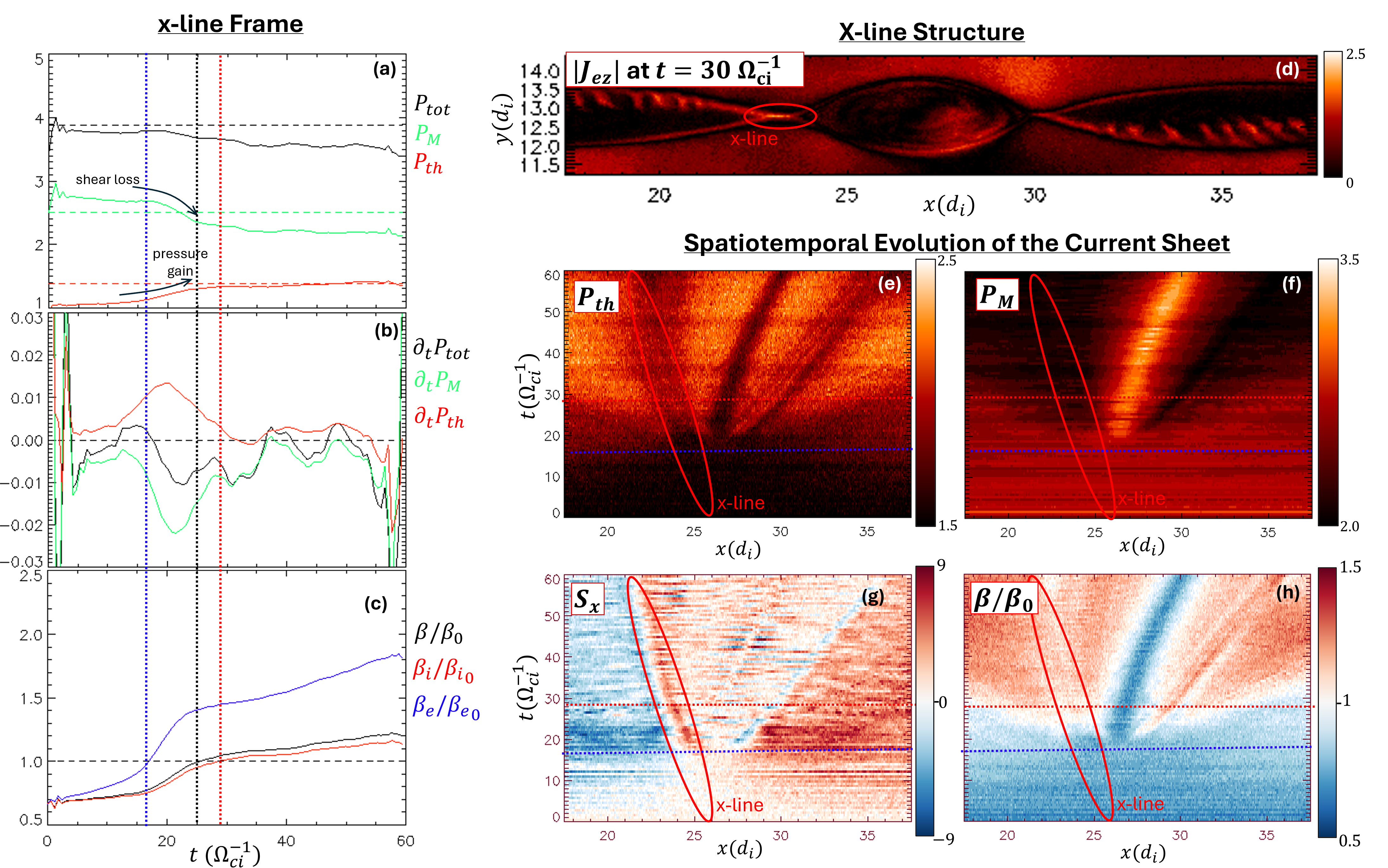}
\caption{The Onset Interval and X-line Pressure Evolution.  (a) $P_{tot}$, $P_M$, and $P_{th}$ in the x-line reference frame (in solid black, green, and red, respectively) with horizontal dashed lines of the same color scheme indicating the initial asymptotic values of their respective variables.  (b) $\partial_t P_{tot}$, $P_M$, and $P_{th}$ in the x-line frame and in black,green,and red, respectively.  (c) $\beta/\beta_0$, $\beta_i/\beta_{i0}$, and $\beta_e/\beta_{e0}$ in the x-line frame, where the ``$0$" subscript indicates the initial asymptotic value. (d) Electron out-of-plane current density $J_{ez}$ of the x-line just after the onset interval, with the primary x-line circled.  (e-h) space vs time plots of the $P_{th}$, $P_M$, outflow component of Poynting flux $S_x$ (used to visually identify the x-line), and the $\beta/\beta_0$ along the PIL, respectively.  The x-line analyzed throughout this study is also circled in panels e-h.} 
\label{onsetinterval}
\end{figure}

The mixed equilibrium achieved early in the simulation is eventually broken by the onset of magnetic reconnection and the formation of an x-line.  Figure \ref{onsetinterval} shows the evolution of various quantities (all smoothed over a $6 \ \Omega^{-1}_{ci}$ interval) at the x-line as the system begins to reconnect.  These include $P_M$, $P_{th}$, and the total pressure ($P_{tot} = P_M + P_{th}$) in figure \ref{onsetinterval}a and their time derivatives in figure \ref{onsetinterval}b.  Figure \ref{onsetinterval}c shows the evolution of $\beta_e$,$\beta_i$, and total $\beta$ each normalized to their respective initial asymptotic values, along with vertical dotted lines indicating when each reaches its initial asymptotic value (when the normalized quantity reaches unity).  Panels c-h in figure \ref{onsetinterval} show the structure of the x-line in time and space.  The reconnection onset process that occurs in figure \ref{onsetinterval} consists of an approximately simultaneous phase of $\partial_t P_M<0$ and $\partial_t P_{th}>0$, mostly occurring after ${\beta_e/\beta_{e0}} = 1$ and before ${\beta/\beta_{0}} = 1$.  There is a net pressure decrease $P_{tot}<0$ due to $|\partial_t P_{th}|<|\partial_tP_M|$. 

\section{\label{sec:level5} Local Thermodynamic Evolution}
\begin{figure}[h]
\centering
\includegraphics[scale=0.55]{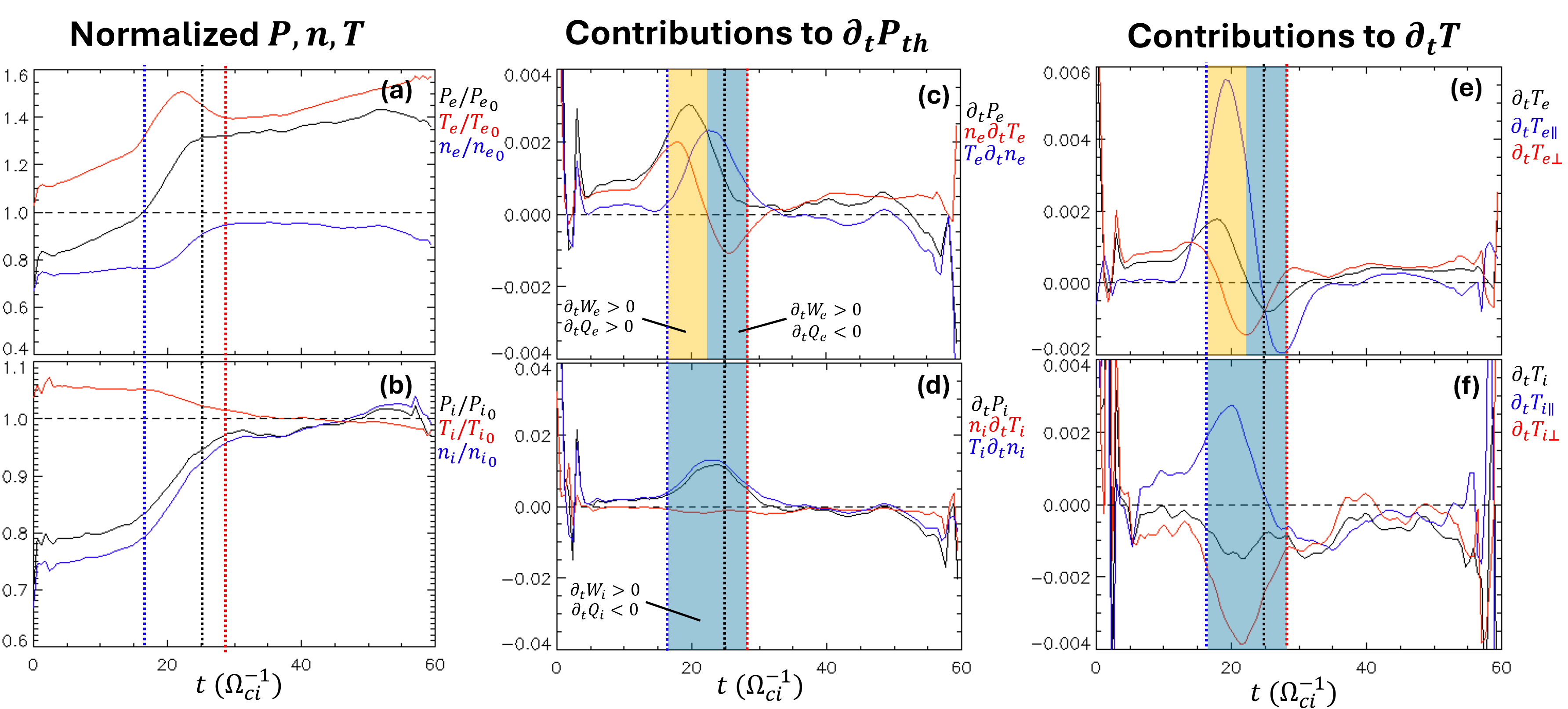}
\caption{Thermodynamic Contributions to Pressure Evolution at the X-line. Panels (a-b) show $P/P_0$, $T/T_0$, and $n/n_0$ at the x-line for electrons (a), and ions (b). Panels (c-d) show $\partial_t P$, $n\partial_t T$, and $T\partial_t n$ for electrons (c), and ions (d).  Panels (e-f) show $\partial_t T$, $\partial_t T_\parallel$, and $\partial_t T_\perp$ for electrons (e) and ions (f).  Regions in yellow and blue indicate intervals of $\partial_t Q>0$ and $\partial_t Q<0$, respectively for electrons (panels c and e) and ions (panels d and f).}
\label{firstlaw}
\end{figure}

Figure \ref{firstlaw}(a-b) shows the evolution of $P_{th}/P_{th0}$, $n/n_0$, and $T/T_0$ for the electrons (figure \ref{firstlaw}a) and ions (figure \ref{firstlaw}b) at the x-line.  The onset begins with increasing $P_e$, $P_i$, $n_e$, $n_i$ and $T_e$, but decreasing $T_i$. The contributions to $\partial_t P_{th}$, $n\partial_tT$ and $T\partial_t n$ by electrons (figure \ref{firstlaw}c) and ions figure \ref{firstlaw}d) suggest that both species undergo compressive work ($T\partial_tn > 0$ and $\partial_t W > 0$) during the onset interval, but the ion contribution is larger by roughly an order of magnitude and accounts for most of the increase in $P_{th}$.  Changes in the ion heat density term ($n_i\partial_tT_i < 0$ and $\partial_t Q_i < 0$) are small compared to the ion compressive work.  In contrast, there is initially an interval of increasing electron heat density ($n_e\partial_tT_e > 0$ and $\partial_t Q_e > 0$) followed by a reversal as the electron compressive work increases.  For the electrons, the heat and work densities have similar magnitudes.  

The contributions to $\partial_t T$ at the x-line are shown for electrons (figure \ref{firstlaw}e) and ions (figure \ref{firstlaw}f), including separate contributions parallel and perpendicular to the local magnetic field.  The $\partial_t T_i < 0$ is largely due to $\partial_t T_{i\perp}<0$, while $\partial_t T_{i\parallel} > 0$.  The $\partial_t T_e > 0$ is due to a significant  $\partial_t T_{e\parallel} > 0$ while $\partial_t T_{e\perp} < 0$.  $\partial_t T_e$ goes through its reversal as the $\partial_t T_{e\parallel}$ and $\partial_t T_{e\perp}$ contributions become negative and comparable in magnitude.

\begin{figure}[h]
\centering
\includegraphics[scale=0.5]{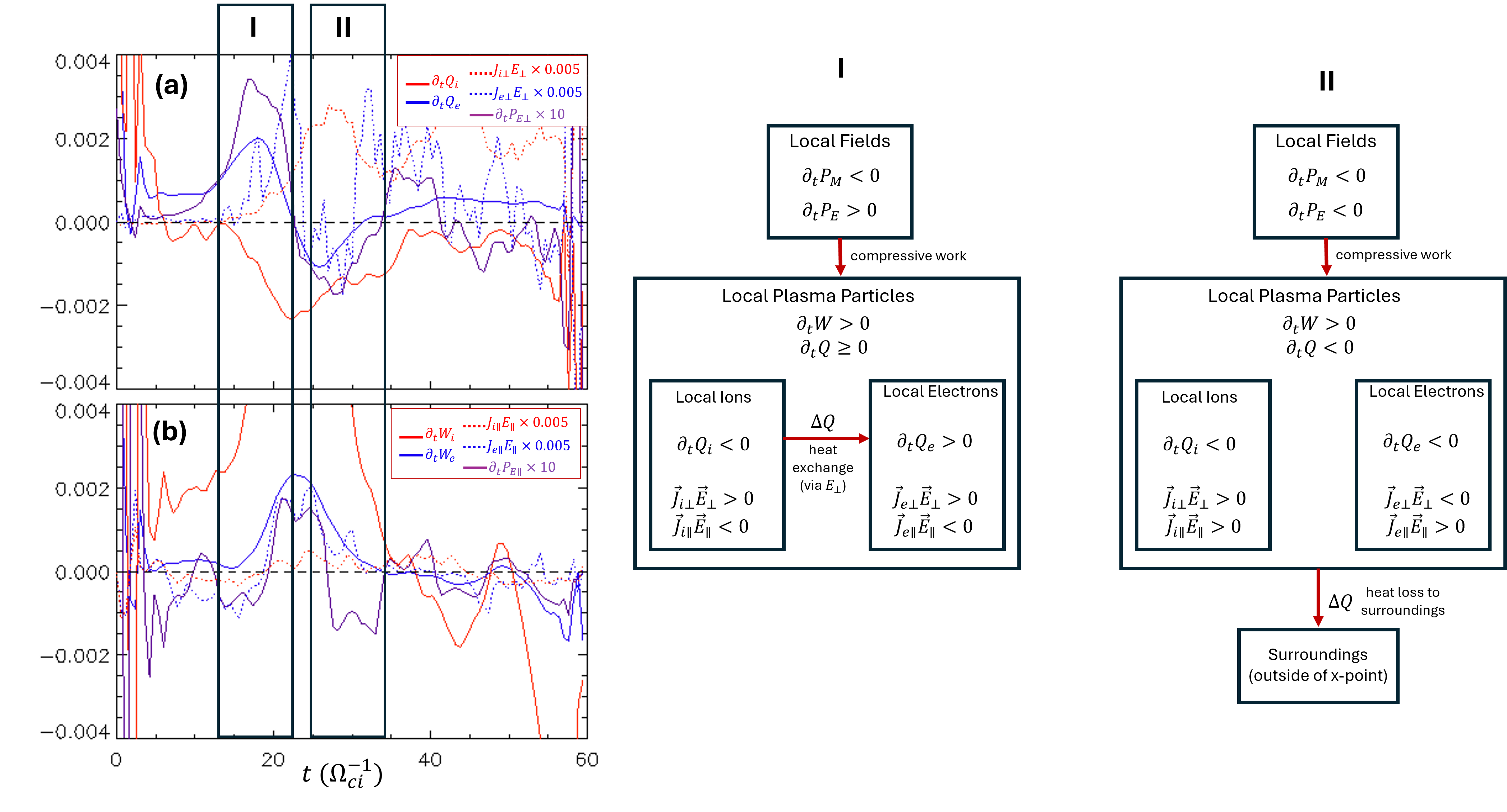}
\caption{Comparison of Power Densities at the x-line.  Panel (a) shows the ion and electron $\partial_tQ$,ion and electron $J_\perp E_\perp$ terms, and the time evolution of the electric field pressure perpendicular to the local magnetic field $\partial_tP_{E\perp}$ .  Panel (b) shows the ion and electron $\partial_tW$ (with the axis range chosen to emphasize $\partial_tW_e$),ion and electron $J_\parallel E_\parallel$ terms, and the time evolution of the electric field pressure parallel to the local magnetic field $\partial_tP_{E\parallel}$.  The diagrams on the right represent the proposed local thermodynamic processes at play in the early and late stages of reconnection onset.}
\label{comparison}
\end{figure}

\section{\label{sec:level7} Discussion}

We have examined the local thermodynamics of an x-line as it forms within a mixed equilibrium current sheet.  The $B_z^2$ enhancement and depletion in $P_{th}$ along the current sheet create a localized $\beta$ depletion while maintaining an approximately uniform $P_{tot}$ across the field reversal.  

During onset, the locally enhanced $P_M$ decreases and the locally depleted $P_{th}$ increases at the x-line, bringing both quantities and the local $\beta$ closer to their asymptotic values (figure \ref{onsetinterval}).  Since $|\partial_t P_M|>|\partial_t P_{th}|$, there is a  net $\partial_t P_{tot}<0$ at the x-line during the early stage of onset.  The equilibration of $\beta$ has contributions from both particle species, with $\beta_e/\beta_{e0}$ approaching unity before $\beta_i/\beta_{i0}$.  The sequence makes it convenient to break the equilibration process into two stages, based on the local value of $\beta$ for each species with respect to upstream parameters:  

\begin{enumerate}

\item $\beta_e/\beta_{e0} \geq 1>\beta/\beta_0$  
\item  $\beta/\beta_{0} \geq 1>\beta_i/\beta_{i0}$ 

\end{enumerate}

The most significant pressure changes occur in the first of these stages, soon after $\beta_e/\beta_{e0} \sim1$.  Overall, $\partial_t P_{th}$ is mainly accounted for by ion compression $T_i \partial_t n_i$ (figure \ref{firstlaw}), which is the largest contribution by almost an order of magnitude.  In the early onset stage,  $\partial_t Q_e > 0$ followed by an interval of $\partial_t Q_e < 0$ of similar magnitude and $\partial_t W_e > 0$.  The $\partial_t Q$ reversal is unique to the electrons while $\partial_t Q_i < 0$ during the entire onset interval, implying that in the early stage of onset the electrons are locally gaining heat density and the ions are losing heat density while in late onset, both species lose heat density $\partial_t Q_{e,i} < 0$  while undergoing compressive work $\partial_t W_{e,i} > 0$.  

The contributions to $\partial_t T$ (figure \ref{firstlaw}(e-f)) reveal that $\partial_t T_i<0$ is mostly due to a large $\partial_t T_{i\perp}<0$ exceeding a simultaneous $\partial_t T_{i\parallel}>0$ while $\partial_t T_e$ and its reversal is mainly due to changes in $T_{e\parallel}$ that tend to offset changes in $T_{e\perp}$.  Near the $\beta/\beta_0 \sim 1$ threshold, the $\partial_t Q_e$ approaches its maximum negative value, indicating that the electron heat loss diminishes after $\beta$ reaches its asymptotic value.     

In figure \ref{comparison}, we again plot some of the relevant power densities, and include contributions from $\partial_t P_E$ and $\vec{J}\cdot\vec{E}$ with scaling factors to ease their comparison with the other terms.  $\partial_t P_E$ is negligible to the overall pressure evolution at the x-line, but its temporal evolution may still provide some insight into the mechanisms of heat transport and compressive work.  Decomposition of the components of $\partial_t P_E$,  $\partial_t Q$ and $\partial_t W$ reveal roughly correlated reversals in $\partial_t Q_e$ and $\partial_t P_{E\perp}$ (shown in figure \ref{comparison}), implying that the buildup and depletion of $P_{E\perp}$ is closely associated with changes in $Q_e$ and $T_e$. $\partial_t P_{E\parallel}$ is somewhat correlated with the largest $\partial_t W$ contributions.  The peak $\partial_t W_{e,i}$ and peak $\partial_t P_{E_\parallel}$ are roughly correlated during the compressive stage ($W_{e,i}$ and $P_{E\parallel}$ are both growing with time).  Similar correlations are seen in $\vec{J}\cdot\vec{E}$ and its components, which are typically orders of magnitude larger than any of the power densities associated with local pressure evolution.

During the buildup phase when the pressure evolution is most significant, the local ions are losing heat density while the electrons are gaining heat density at a comparable rate $-\partial_t Q_i \sim \partial_t Q_e$ (interval I in figure \ref{comparison}), and the local $P_E$ is increasing largely due to the $P_{E\perp}$ component.  The near-simultaneous gain in $Q_e$ and loss of $Q_i$ is particularly interesting.  While their correlation alone does not definitively determine a direct relationship between the ion heat loss and the electron heat gain, the gain in $P_E$ and in particular $P_{E\perp}$ during this interval suggests that they may be related, as ion heat may be transferred to the electrons via $E_\perp$ as an intermediary step $Q_i \rightarrow P_{E_\perp}\rightarrow Q_e$.  The growth of $J_\perp E_\perp$ for both species also begins in this stage, suggesting that this accumulating electric field energy facilitates electromagnetic energy transfer into the local particles.  While $J_{i\perp} E_\perp$ continues to increase (interval II in figure \ref{comparison}), $J_{e\perp} E_\perp$ reverses sign close to the  $\partial_tP_{E\perp}$ and $\partial_t Q_e$ reversal, again suggesting a mechanistic relationship between electron energy gain/loss normal to the magnetic fields, normal electric field evolution, and changes in local electron heat density.  Meanwhile, the work density evolution of both species $\partial_t W_{i,e}>0$ seems to be more correlated with $\partial_t P_{E\parallel}$ and $J_{e\parallel}E_\parallel$, at least in the early stage of the onset process.  These correlations suggest that in the large guide field scenario, the compressive work associated with local magnetic shear loss is accompanied by time-evolving parallel electric fields and parallel particle energization. This result is somewhat consistent with recent analysis \cite{payne2025situ} of an electron-only reconnection event observed by MMS, which also found evidence of field-aligned particle energization associated with compressive work due to local magnetic curvature changes.

\section{\label{sec:level8} Conclusion}
In this study, we have examined the process of magnetic shear equilibration, particularly how the evolution of local magnetic shear during reconnection onset influences the local evolution of thermal pressure via changes in local heat and work density.  The loss of excess magnetic shear coincides with a gain in thermal pressure that counteracts but does not fully balance out the magnetic pressure loss, leading to a slight reduction in total pressure as $\beta$ at the current sheet approaches its asymptotic value.  The thermal pressure evolution comes from a combination of time-evolving temperature and time-evolving density.  The thermal pressure evolution is largely dominated by ion density evolution term $T_i \partial n_i$, which we argue is synonymous with the statement that that local ion work density $W_i$ (work on the local ions per unit volume) accounts for most of the time evolution of the local internal energy density associated with the plasma thermal energy.  Although the ion evolution accounts for the particle energy budget, the evolution of the electron pressure has comparable contributions from work and heat density evolution, which tend to be more correlated with the local changes in the parallel and perpendicular electric fields, respectively.

\section*{Data Availability}
The simulation data used in this study is available upon request from the authors

\section*{Acknowledgements}
The primary funding for this work comes from the NASA MMS Project through the MMS Early Career Grant 80NSSC23K1662.
\bibliography{Shear}

\end{document}